\documentclass[11pt]{article}

\usepackage{graphicx}

\usepackage{color}

\usepackage{amsmath, amssymb,amsfonts,longtable,tabularx}
\usepackage{cite}
\usepackage{colortbl}
\usepackage[all]{xy}
\usepackage{epsfig}
\usepackage{subfigure}

\usepackage{setspace}
\usepackage{caption}
\usepackage{pifont}
\usepackage[table]{xcolor}

\definecolor{gray1}{rgb}{0.9,0.9,0.9}
\definecolor{gray2}{rgb}{0.8,0.8,0.8}

\usepackage[hmargin=2cm, vmargin=2cm]{geometry}

\date{}

\usepackage{cite}

\begin{document}

\begin{flushleft}
{\huge
\textbf{Under-dominance constrains the evolution \\ of negative autoregulation in diploids}
}
\bigskip
\\
Alexander J.Stewart$^{1\ast}$, 
Robert M. Seymour$^{1,2}$, 
Andrew Pomiankowski$^{1,3}$ \&
Max Reuter$^{3}$
\\
\bigskip
\bigskip
$^1$ CoMPLEX, University College London, London, UK
\\
$^2$ Department of Mathematics, University College London, London, UK
\\
$^3$ Department of Genetics, Evolution and Environment, University College London, London, UK
\\
$\ast$ E-mail: alstew@sas.upenn.edu. Present address: Department of Biology, University of Pennsylvania, Philadelphia, PA, USA
\end{flushleft}

\vspace{1cm}

\begin{abstract} 
 Regulatory networks have evolved to allow gene expression to rapidly track changes in the environment as well as to buffer perturbations and maintain cellular homeostasis in the absence of change \cite{ABC:Eldar,Raj:2008bs,Wang:2011}. 
Theoretical work and empirical investigation in \textit{Escherichia coli} have shown that negative autoregulation confers both rapid response times and reduced intrinsic noise \cite{Becskei:2000vo,ABC:Lestas,Rosenfeld:2002iv,Thattai:2001qp}, which is reflected in the fact that almost half of \textit{Escherichia coli} transcription factors are negatively autoregulated \cite{ABC:She2002,Thieffry:1998et,Warnecke:2009fi}. However, negative autoregulation is exceedingly rare amongst the transcription factors of \textit{Saccharomyces cerevisiae} \cite{ABC:Gue2002,ABC:Lee2002,Milo:2004ez,Thieffry:1998et,Warnecke:2009fi}. This difference is all the more surprising because \textit{E.~coli} and \textit{S.~cerevisiae} otherwise have remarkably similar profiles of network motifs \cite{Milo:2004ez}. In this study we first show that regulatory interactions amongst the transcription factors of  \textit{Drosophila melanogaster} and humans have a similar dearth of negative autoregulation to that seen  in \textit{S.~cerevisiae}. We then present a model demonstrating that this fundamental difference in the noise reduction strategies used amongst species can be explained by constraints on the evolution of negative autoregulation in diploids. We show that regulatory interactions between pairs of homologous genes within the same cell can lead to under-dominance --- mutations which result in stronger autoregulation, and decrease noise in homozygotes, paradoxically can cause increased noise in heterozygotes. 
This severely limits a diploid's ability to evolve negative autoregulation as a noise reduction mechanism. 
Our work offers a simple and general explanation for a previously unexplained difference between the regulatory architectures of \textit{E. coli} and yeast, \textit{Drosophila} and humans.
It also demonstrates that the effects of diploidy in gene networks can have counter-intuitive consequences that may profoundly influence the course of evolution.
\end{abstract}

\section*{Author Summary}
All genes have to deal with intrinsic noise, and a variety of mechanisms have evolved to reduce it.
One important mechanism of noise reduction for transcription factors is negative autoregulation, in which a gene product represses its own rate of transcription.
Negative auotregulation occurs frequently in \textit{E. coli} but, we find, occurs extremely rarely in \textit{S.~cerevisiae},  \textit{D. melanogaster} and humans.
Whilst there are a great many important differences in the genetic architectures of these organisms, they tend to share, with the exception of negative autoregulation, similar profiles of network motifs.
This makes the discrepancy in the degree of negative autoregulation all the more striking, as it lacks any obvious explanation.
Our study presents a potential explanation, by comparing the evolvability of negative autoregulation as a noise reduction mechanism in haploids and diploids.
We show that, in diploids, mutations that increase the strength of negative autoregulation at one gene copy often increase overall noise in gene expression.
This results in under-dominance, in which heterozygotes are less fit than homozygotes.
The result is that the evolution of negative autoregulation in diploids is significantly constrained.
We verify our results using a combination of detailed molecular simulations and evolutionary simulations

\section*{Introduction}

\noindent 
Negative autoregulation is a network motif in which a transcription factor inhibits its own expression. Theoretical work has shown that this type of regulation reduces intrinsic noise and quickens the response time to environmental perturbations \cite{Becskei:2000vo,Rosenfeld:2002iv,Thattai:2001qp} and experiments using artificial gene regulatory circuits in \textit{E.~coli} have confirmed these predictions \cite{Rosenfeld:2002iv}. Negative autoregulation therefore represents a simple yet powerful mechanism to maintain cellular homeostasis in the face of environmental and metabolic perturbations and reduce the often substantial fitness costs that noise can incur \cite{Wang:2011}. Different organisms, however, vary a great deal in their use of the motif. In \textit{E.~coli}, close to 50\% of transcription factors (82 out of 182) \cite{ABC:She2002,Thieffry:1998et,Sanchez:2008oa,Warnecke:2009fi} have been shown to negatively autoregulate. In contrast, negative autoregulation is almost entirely absent from the transcription factors of  \textit{S.~cerevisiae} (3 out of 169) \cite{ABC:Gue2002,ABC:Lee2002,Milo:2004ez,Thieffry:1998et,Warnecke:2009fi}.

How can we account for this striking discrepancy? In order to answer this, we looked at the extent to which negative autoregulation is used in other species. We interrogated systematic datasets on the regulatory interactions amongst the transcription factors of  \textit{D.~melanogaster} and humans and found a similar pattern to that observed in yeast: in \textit{D.~melanogaster} 3 out of 87 \cite{Bergman:2005vn,Matys:2006fk,Wingender:1996uq} and in humans 5 out of 301 \cite{Bauer:2011kx,Matys:2006fk,Wingender:1996uq} transcription factors autoregulate (see SI, Table S1-S3). Currently, there is no obvious way to account for this striking discrepancy between these organisms, despite widespread interest in the strategies they employ to tackle noise \cite{ABC:Eldar,Raj:2008bs,Wang:2011,Becskei:2000vo,ABC:Lestas,Rosenfeld:2002iv,Thattai:2001qp}. Here we develop a model, founded in biophysics, for the evolution of negative autoregulation in diploid species. We use it to argue that a dearth of autoregulating genes in yeast, flies and humans can be explained by constraints on the evolution of negative autoregulation that arise due to diploidy.  
\section*{Results}

\subsection*{Gene expression under negative autoregulation} Previous theoretical work on the dynamics of gene expression under negative autoregulation has considered single genes and so is implicitly haploid \cite{Becskei:2000vo,ABC:Lestas,Rosenfeld:2002iv,Thattai:2001qp}. Such models exclude the more complex interactions that occur due to cross-regulation between homologous gene copies within a diploid cell (Fig.~1). Here we characterise the expression dynamics and regulatory evolution of homologous pairs of negatively autoregulating  genes, taking into account the cross-talk between alleles. 

\begin{figure}[h!]
        \centering
          \subfigure[]{
\includegraphics[scale=0.4]{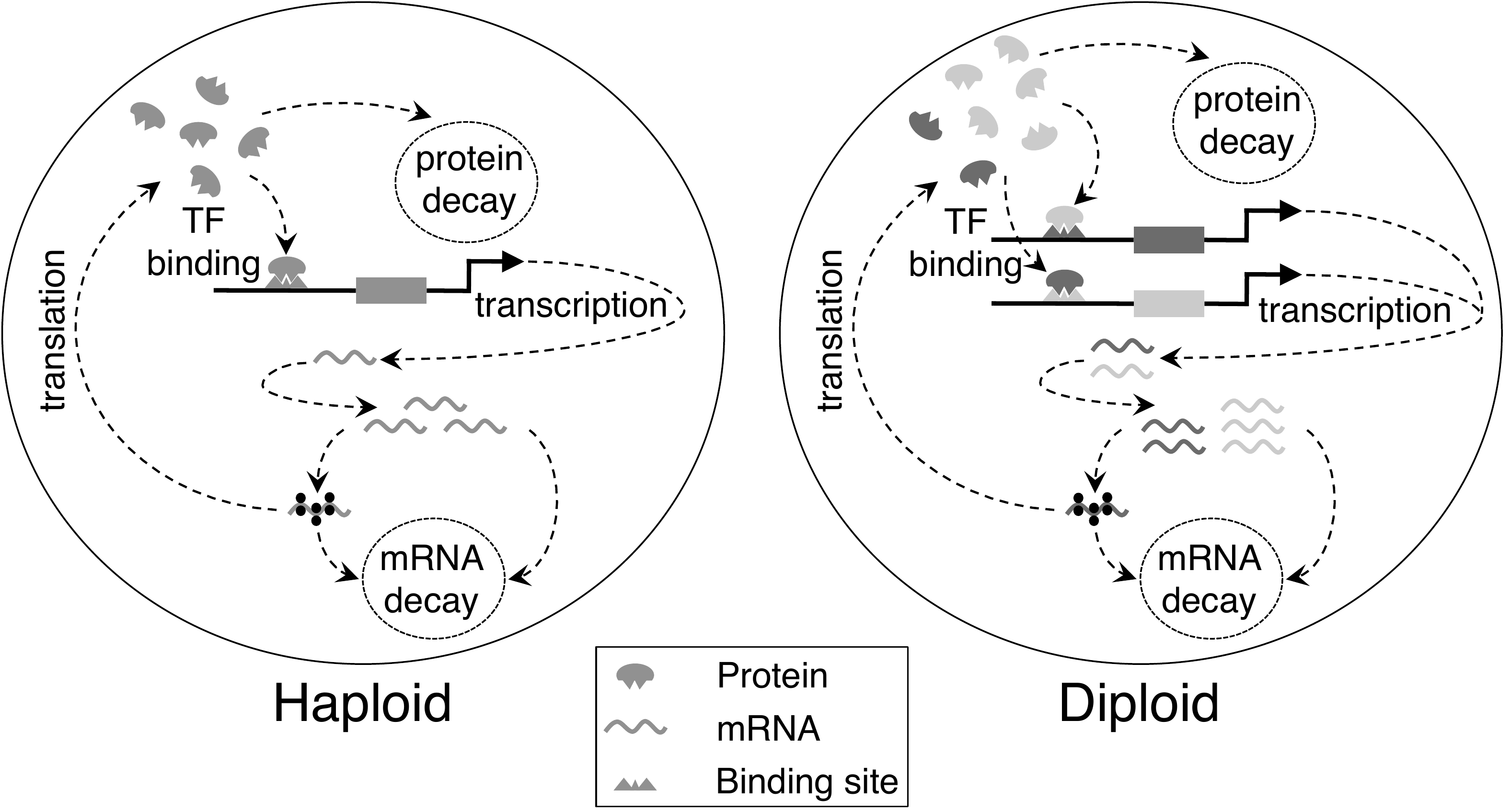} 
}
  \subfigure[]{
  \includegraphics[scale=0.4]{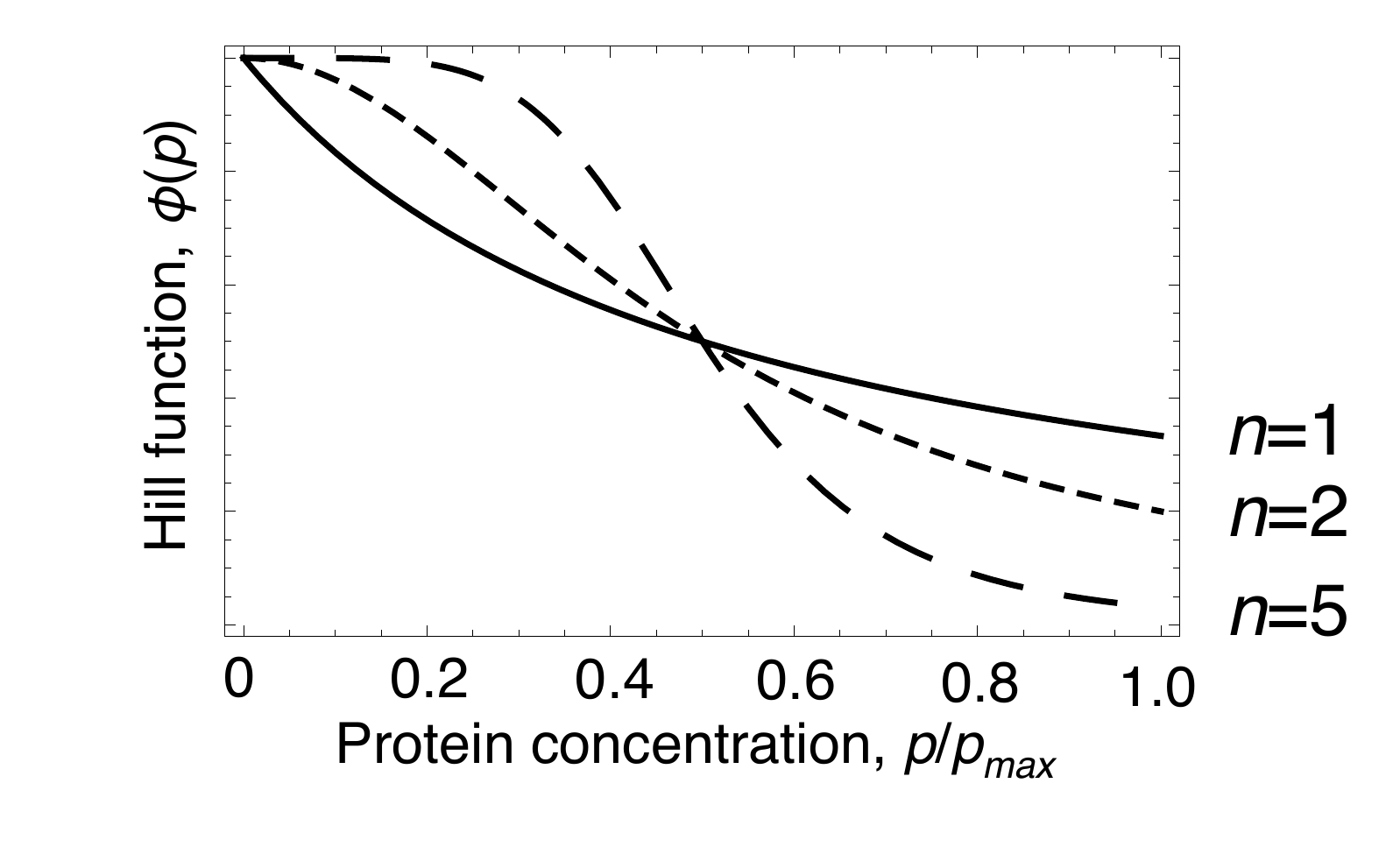} 
  }
\caption{\textbf{Cross-talk in diploid autoregulators}\ \  
(a) Schematic representation of negative autoregulation when one (left) and two (right) copies of a gene are present in a cell. In the haploid the amount of negative autoregulation the gene experiences depends on on its own expression level. In the diploid, two gene copies are present (shown as light gray and dark gray), and the amount of negative autoregulation experienced by each gene depends on the expression level of \emph{both} genes combined. If the two gene copies differ from one another in the strength of their transcription factor binding sites, complex dynamics can arise that are not observed in haploids. (b) IIllustration of variation in the repression function, $\phi(p)$, with protein concentration for different Hill coefficients, $n=1$ (sold line), $n=2$ (small dashes) and $n=5$ (large dashes).}
\end{figure}

\clearpage

We model negative autoregulation in a diploid using a set of ordinary differential equations that track changes in the mRNA and protein concentrations for each of a pair of alleles (labelled with subscripts $i$ and $j$), $r_i$, $r_j$, $p_i$ and $p_j$. The total concentration of mRNA and protein in the diploid cell are given by the summed output of the two alleles $r=r_{i}+r_{j}$ and $p=p_{i}+p_{j}$. Changes in mRNA and protein concentrations for the pair of alleles over time are given by
\begin{eqnarray}
\label{ODE}
\nonumber \frac{dr_{i}}{dt}&=&k_l+\phi_{i}(p)-\gamma_{r}r_{i},\\
\nonumber \frac{dr_{j}}{dt}&=&k_l+\phi_{j}(p)-\gamma_{r}r_{j},\\
\nonumber \frac{dp_{i}}{dt}&=&r_{i}k_{p}-\gamma_{p}p_{i},\\
\frac{dp_{j}}{dt}&=&r_{j}k_{p}-\gamma_{p}p_{j}.
\end{eqnarray}
\\
According to these equations, mRNA is transcribed at a (usually low) constant background rate $k_l$, plus a rate $\phi(p)$ due to negative autoregulation, that decreases as the total cellular protein level $p=p_{i}+p_{j}$ increases. Protein is produced from mRNA at the rate of translation $k_{p}$, whilst protein and mRNA degrade with rates $\gamma_{p}$ and $\gamma_{r}$, respectively.

As in previous work \cite{Becskei:2000vo,Rosenfeld:2002iv}, we model the repression function $\phi(p)$ in Eqs.~\ref{ODE} as a Hill function
\begin{equation*}
\phi_{i}(p)=\frac{k_{0}}{1+\left(\frac{p}{K_{i}}\right)^n}
\end{equation*}
\\
where $K_i$ is the dissociation constant associated with the autoregulating transcription factor binding site. Smaller values of $K$ (lower rates of dissociation) indicate a steeper slope and stronger regulation. The Hill coefficient $n$ governs the steepness of the function at the inflection point and hence determines how step-like regulation will be. In systems where transcription is regulated by a single binding site, $\phi(p)$ has a Michaelis-Menten-like form, corresponding to a Hill coefficient of $n=1$ \cite{Chu:2009vg,Gerland:2002cv,Rosenfeld:2002iv}. A single binding site is the simplest and most relevant case for evolving negative autoregulating, and it is the one we focus on here. For completeness, we analyse the more general case of arbitrary Hill coefficient in the Appendix and in the Supporting Information we show that our results also hold for different values of $n$.

In the absence of negative autoregulation (i.e., $\phi(p)=k_{0}$), mRNA is produced at the maximum rate of transcription  $k_l+k_{0}$. In this case, concentrations of mRNA and protein reach equilibrium values of $r_{max}=\frac{2(k_{0}+k_{l})}{\gamma_{r}}$ and $p_{max}=r_{max}\frac{k_{p}}{\gamma_{p}}$. Starting from these values, equilibrium mRNA and protein levels decrease with increasing autoregulatory binding strength (decreasing $K$). The minimum mRNA and protein levels are reached when negative autoregulation is strongest (i.e. $\phi(p)\to 0$ as $K\to0$). The resulting minimum equilibrium concentrations are $r_{min}=\frac{2k_{l}}{\gamma_{r}}$ and $p_{min}=r_{min}\frac{k_{p}}{\gamma_{p}}$. 

\subsection*{Evolution of negative autoregulation for homeostasis and faster response times} 
In order to analyse the evolution of autoregulatory binding sites we consider two separate but related functions of negative autoregulation: faster response times and maintaining mRNA and protein homeostasis.
First, to study the evolution of negative autoregulation for faster response times, we simply equate the fitness of a system with its response time (i.e the time taken to return to equilibrium following a perturbation).
We use Eqs.~1 to infer selection pressures on the strength of autoregulation, i.e., the dissociation constant $K$, by analysing how quickly genotypes with different autoregulatory binding strength return to equilibrium following a perturbation in protein level. To do this we calculate a genotype's ``response time'': the time taken for cellular protein concentration to return to equilibrium following a perturbation. We model perturbations as a reduction of the protein level to a fraction $\alpha$ of the equilibrium level. The value of $\alpha$ varies continuously between $0$ and $1$ to encompass both small perturbations, for example those resulting from intrinsic noise in transcription and translation ($\alpha\approx1$), and larger perturbations, for example those resulting from resource deprivation in the environment or following cell division \cite{Raj:2008bs}. We present results derived from numerical analysis of Eqs.~\ref{ODE} that are applicable to perturbations of any size. These are complemented with an analytical treatment of the response time of the system to small perturbations, based on its maximal eigenvalue, $\left|\lambda\right|$ (see Appendix), which allows us to develop an intuition for how autoregulating genes in diploids respond to perturbations.

To study the evolution of negative autoregulation for homeostasis, we turn to stochastic simulations of negatively autoregulating genes, which allow us to assess the amount of intrinsic noise associated with gene expression.
Previous work has shown that negative autoregulation can help maintain homeostasis in gene expression by reducing the amount of intrinsic noise in negatively autoregulating genes, compared to other genes \cite{Thattai:2001qp}. In fact, reducing the response time of a gene to very small perturbations away from equilibrium, also decreases the intrinsic noise in gene expression.
Therefore, the two functions of negative autoregulation we consider (producing faster response times and reduced intrinsic noise) are highly inter-related.
To study the evolution of negative autoregulation for lower intrinsic noise, we equate the fitness of the system with the amount of intrinsic noise it displays (i.e the ratio of the variance in gene expression to the mean gene expression level).
We infer selection pressures on the strength of autoregulation, i.e., the dissociation constant $K$, by the intrinsic noise of genotypes with different autoregulatory binding strengths.
These are determined by performing Monte Carlo simulations for a full, molecular model of transcription, translation and autoregulation (see Materials and Methods).

\subsection*{Response time in homozygotes} 
We first compare the response times of two homogozyotes whose alleles are identical in every respect except for the dissociation constant. One homozygote carries two copies of a resident allele with dissociation constant $K_1$, the other carries two mutant alleles that have a decreased dissociation constant $K_2=K_1 \exp[-\epsilon]$ (with $\epsilon>0$) and hence stronger autoregulatory binding. Numerical analysis of the system shows that homozygotes for the more strongly autoregulating allele (with $K_{2}$) respond more quickly than homozygotes for the more weakly autoregulating allele (with $K_{1}$, Fig.~2a). This is true up to a value of $K\approx K_{opt}$, which provides the fastest response time attainable by the system and hence provides the optimal binding strength. Further increases in regulation beyond this value are not favoured because they lead to overshooting the optimal binding strength. These results for diploid homozygotes mirror those obtained for haploids \cite{Rosenfeld:2002iv} (see Appendix) and show that regulatory interactions between pairs of identical alleles do not, in themselves, diminish the beneficial effects of negative autoregulation. Negative autoregulation can therefore, in principle, function as a mechanism to produce faster response times in diploids just as it does in haploids.

\begin{figure}[h!]
       \centering
                \subfigure[]{
\includegraphics[scale=0.35]{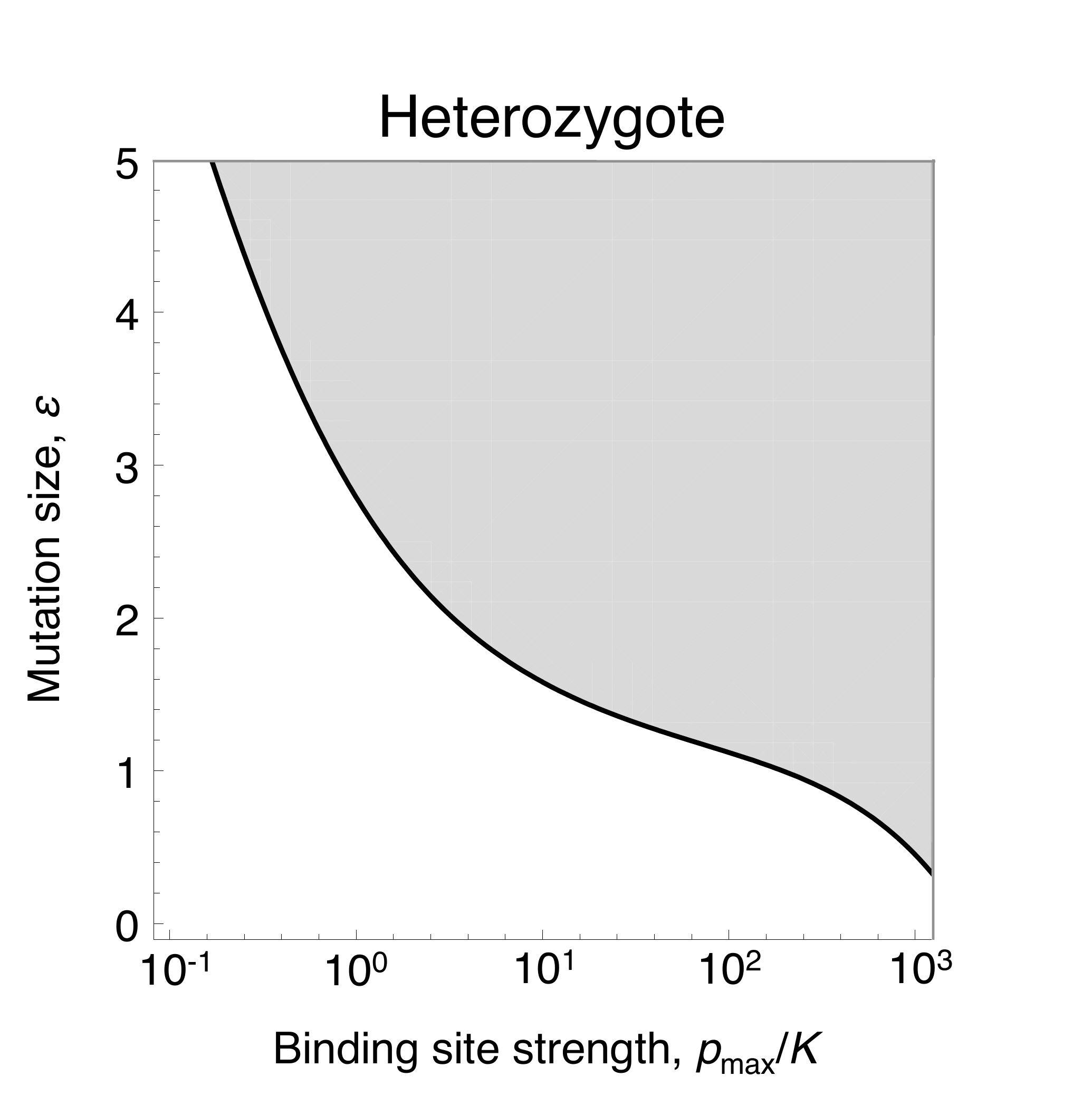} 
\label{fig:subfig1} 
} 
\subfigure[]{
\includegraphics[scale=0.35]{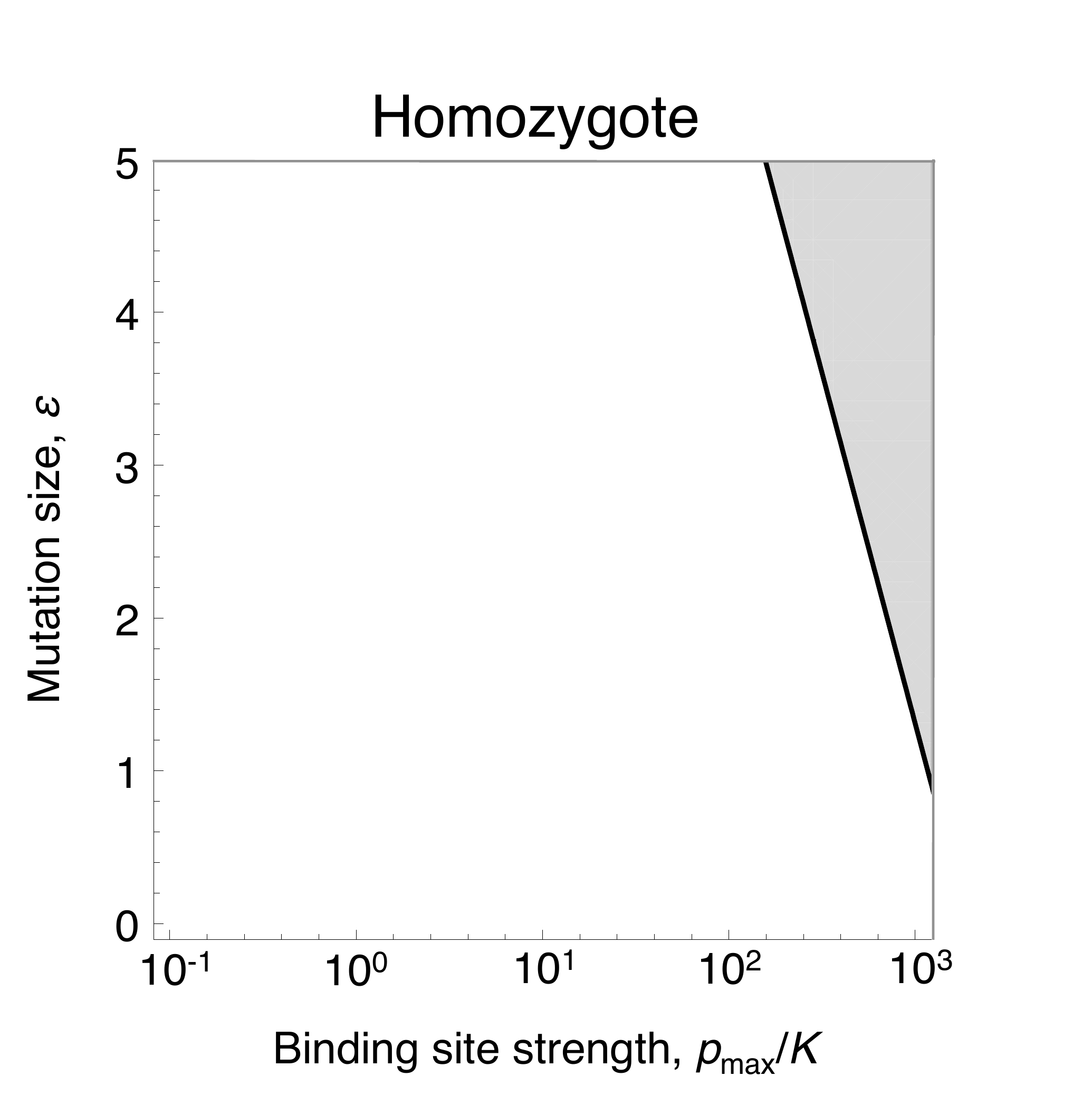}
\label{fig:subfig1} 
} 
\caption{\textbf{Invasibility of autoregulatory binding sites}\ \ 
The response time of mutant (a) homozygotes and (b) heterozygotes are shown. Different values of the binding strength of the resident allele, in units of $p_{max}/K$ (x-axis), are plotted against mutations to binding site strength $\epsilon$ of different size (y-axis). Thus the graphs compare a resident allele, $K_{1}$ with a mutant allele, $K_{2}=K_1\exp[-\epsilon]$. Mutations falling into white region result in decreased response time in the carrier compared to resident genotype and are favoured by selection; mutations falling into the gray region result in increased response time and are not favoured by selection. Weak binding occurs when $p_{max}/K\gtrsim 10^0$ [4,6]. Response times were calculated by numerically integrating Eq.~1 from zero protein concentration to 90\% of the equilibrium. The optimal binding strength in these graphs is $p_{max}/K=1250$, corresponding to a background transcription rate $k_{l}/\gamma_p=10^{-3}$.}
	\label{fig:effect.mu.t}
\end{figure}

\clearpage

\subsection*{Response time in heterozygotes} 
The results above depend on comparing homozygotes for alleles with different dissociation constants, $K_1$ and $K_2$. The evolution of negative autoregulation, however, must occur through the stepwise accumulation of new mutations that are initially rare and found only in heterozygotes. In order to assess whether autoregulation can evolve in diploids, we therefore need to determine whether a mutant allele with a stronger binding site ($K_{2}$) will confer a selective advantage to a heterozygote that also carries a resident allele with a weaker binding site ($K_{1}$). A mutation will be favoured and increase in frequency if a heterozygote is able to respond more quickly to perturbations than a homozygote carrying two copies of the more weakly binding resident allele. 

Numerical analysis of Eqs.~1 reveals that heterozygotes often have greater response times than homozygotes with the more weakly binding resident allele. Fig.~2b shows that heterozygotes only have improved response times when the resident allele binding strength is weak ($K\gtrsim p_{max}$ \cite{Becskei:2000vo,Rosenfeld:2002iv}) or if the effect of a mutation that increases binding strength is small ($\epsilon$ is small). As the resident allele binding strength increases (i.e. $p_{max}/K$ increases) an ever larger range of mutation sizes result in increased heterozygote response times (Fig.~2b), resulting in under-dominance (i.e. heterozygote disadvantage). Typicaly mutation sizes $\epsilon$ for transcription factor binding sites are in the range $1<\epsilon<3$, \cite{Gerland:2002cv,Lassig:2007fk,Berg:2004uq}. In this range regulatory mutations are subject to under-dominance and increasingly so as the binding strength of the resident allele increases. As a consequence, the maximum binding strength that can evolve is likely to be significantly lower than in haploids (Fig.~2). Based on these results, we expect under-dominance to pose a significant barrier to the evolution of negative autoregulation in diploids. 

To better understand why under-dominance arises in this system, we calculated the eigenvalues associated with Eqs.~1. These provide a measure of the rate at which the system returns to equilibrium following a small perturbation, and allow us to elucidate the relative contributions of the different alleles to the response dynamics of the gene pair. The maximal eigenvalue of Eqs.~1 for a heterozygote, $\left|\lambda_{het}\right|$, can be expressed as

\begin{equation}
\label{lambda_het}
\left|\lambda_{het}\right|=\left|\lambda_{hom}\right|-\frac{V}{p_{het}^*},
\end{equation}
\\
(see Appendix) where $V$ is the squared difference of the mean steady state expression levels of the two alleles in the heterozygote and $\left|\lambda_{hom}\right|$ is the maximal eigenvalue of a homozygote with protein concentration equal to that of the heterozygote at equilibrium, $p^*_{het}$ (see Appendix). Eq.~2 says that, even if increasing autoregulatory binding strength leads to a faster response time in a homozygote, this advantage is offset in the heterozygote by an amount $V/p^*_{het}$, which measures how different the expression levels of the two alleles are  (it is analogous to the Fano factor, a measure of the spread in a probability distribution \cite{Thattai:2001qp}). As the difference in the expression of the alleles increases,  $V/p^*_{het}$ increases from $0$ to a maximum $p^*_{het}/2$. 

We can understand why increasing the difference in allelic expression results in increased response time by considering the contribution of the individual alleles to the response time of the gene pair (Fig.~3). The level of negative autoregulation at each allele depends on the strength of its binding site and the amount of protein product present in the cell. In a heterozygote, the allele with the stronger binding site is more strongly suppressed (compared to the same allele in a homozygote), since there is more protein available to bind to it. At the same time, the allele with the weaker binding site is less strongly suppressed compared to the same allele in a homozygote. As a result, the allele with the stronger binding site has a faster response time than in a homozygote, whilst the allele with the weaker binding site has a slower response time than in a homozygote. However, the overall effect tends to be to increase the response time of the heterozygote, because the dynamics of protein expression in the heterozygote are dominated by the allele with the weaker binding site (Fig.~3).

\begin{figure}[h!]
        \centering
               \subfigure[]{
\includegraphics[scale=0.5]{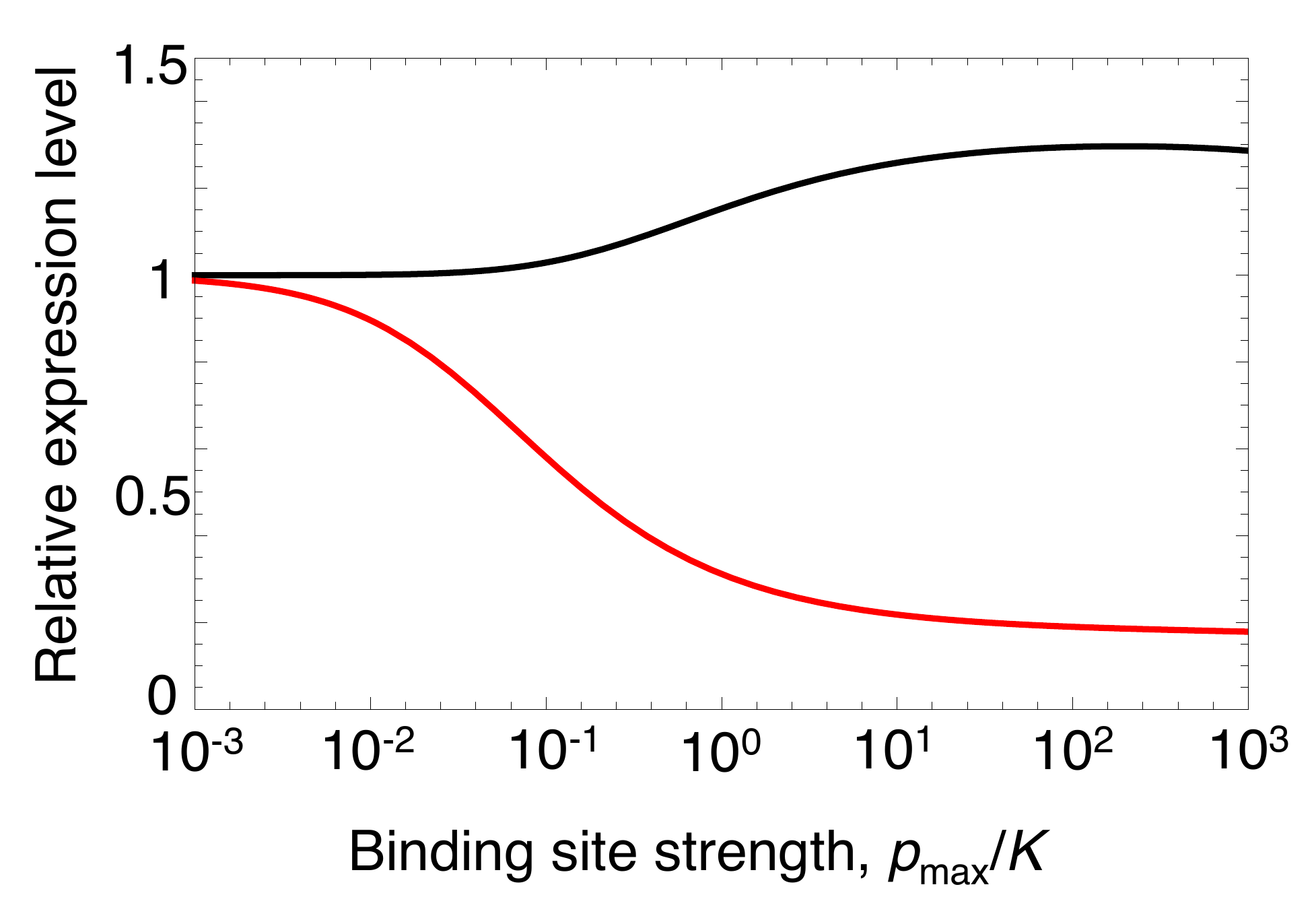} 
\label{fig:subfig1} 
} 
\subfigure[]{
\includegraphics[scale=0.5]{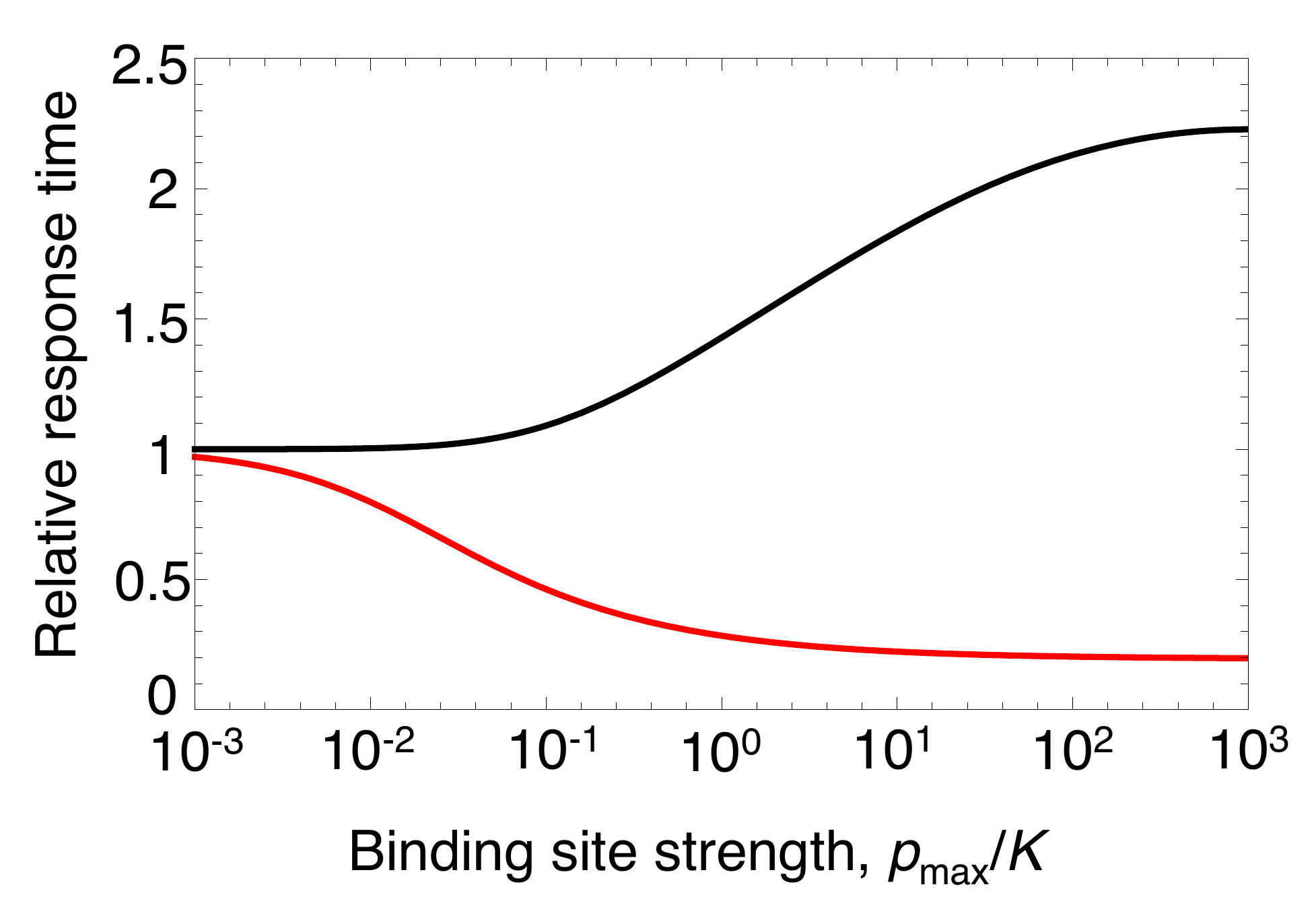}
\label{fig:subfig1} 
}
	\caption{\textbf{Response times and allele expression}\ \ 
	This figure shows quantitative results for the contributions of different alleles to expression and to response time. (a) Expression level of the resident allele (black line) and the mutant allele (red line) in the heterozygote relative to the resident allele in the homozygote. As binding strength increases the resident allele is over-expressed. (b) Response times for individual alleles (time to return to $90\%$ of the equilibrium expression level) in the heterozygote. The response time of the resident allele (black line) and the mutant allele (red line) in the heterozygote are shown relative to the response time of the resident allele in the homozygote. The resident allele in the heterozygote shows an increased response time with increasing binding strength. Mutant alleles in these graphs have dissociation constant $K\exp[-2]$, and the optimal binding strength in these graphs is $p_{max}/K=1250$, corresponding to a background transcription rate $k_{l}/\gamma_p=10^{-3}$.}
\end{figure}

\clearpage

\subsection*{Evolution of faster response times} 
Under-dominance for response time occurs across a wide range of parameter values, but can be avoided if mutations have small effects on binding site strength (Fig.~2b). To determine whether a series of mutations with small effect could offer a feasible way for genes to evolve strong negative autoregulation in diploids, we carried out simulations of binding site evolution that incorporated established properties of real binding sites.

Transcription factor binding sites in eukaryotes vary between $5$ and $\sim30$ nucleotides in length, with an average of 10 nucleotides \cite{Bryne:2008fk}. They have a small number of optimal sequences that bind the transcription factor with maximum affinity \cite{ABC:Har2004,Gerland:2002cv,Gerland:2009nh,Lassig:2007fk,Berg:2004uq}. The binding strength of a site can be expressed as a function of the total binding energy $E$ of its sequence, so $K=\exp[-E]$. This total binding energy is generated by the additive contributions of individual nucleotides to overall binding,  $E=\sum_i \epsilon_i$. Individual contributions are set to $\epsilon_i= 0$  for nucleotides that do not match the optimal sequence and $\epsilon_i>0$ for matched nucleotides  \cite{Gerland:2002cv,Lassig:2007fk,Berg:2004uq}. 

Based on these properties, we performed simulations of the evolution of an autoregulatory binding site under selection for decreased response time. These took into account the empirical distribution of binding site length in model eukaryotes and the variation in contributions to binding strength $\epsilon_i$ across the binding site sequence (see Materials and Methods). The values of $\epsilon_i$ were drawn from a uniform distribution in the interval $(0,3)$. This sampling covers the empirically estimated range $1<\epsilon_i<3$ \cite{Gerland:2002cv,Lassig:2007fk,Berg:2004uq}. It also ensures that mutations of small effect ($\epsilon<1$) occur frequently and so allows for the possibility that autoregulation could evolve via the accumulation of mutations with small effect. Evolution was started from a state of minimum affinity (all nucleotides non-optimal) and proceeded through a series of single nucleotide substitutions. A mutant was assumed to go to fixation if it resulted in a response time less than or equal to that of the resident. Simulations were carried out for both haploids and diploids (for which the response time of mutants was evaluated in the heterozygote state).

The results (Fig.~4) confirm that under-dominance strongly constrains the evolution of negative autoregulation in diploids. Haploids readily evolved binding sites with dissociation constants close to $K_{opt}$. In contrast, the average binding strength in diploids was around 100 times weaker than $K_{opt}$ and only a small proportion of sites reached binding strengths comparable to those of haploids. This shows that under realistic conditions, diploids will rarely be able to evolve the level of autoregulation observed in haploids.

\begin{figure}[h!]
        \centering
\includegraphics[scale=0.5]{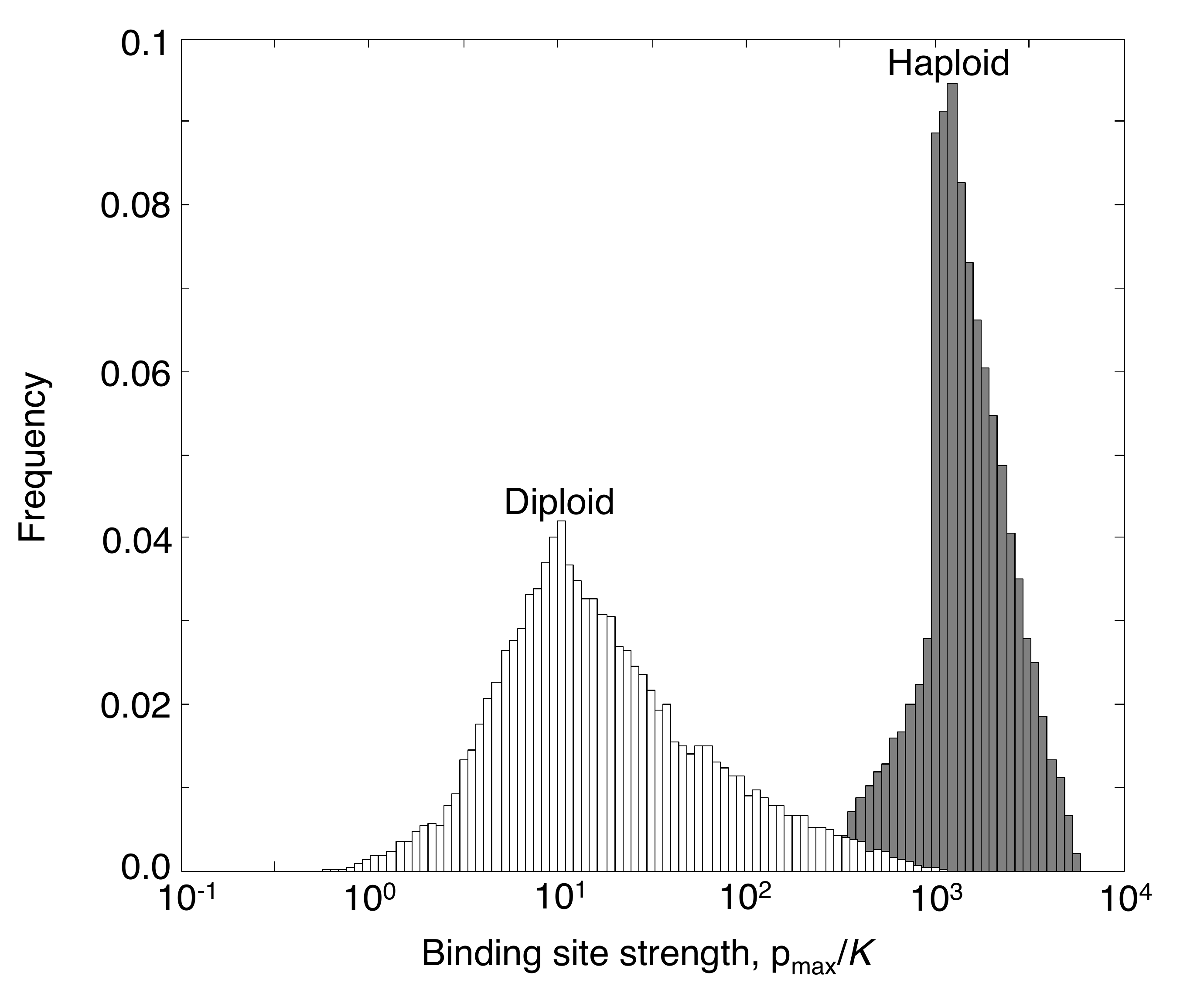} 
\caption{\textbf{Evolution of autoregulatory binding sites}\ \  
Distribution of binding site strength achieved in evolutionary simulations for haploids (gray) and diploids (white). Haploids are able to evolve stronger binding than diploids. The histograms shows results of $10^5$ replicate simulations for each ploidy level. The simulation procedure is described in the main text and the Materials and Methods. The optimal binding strength used was $p_{max}/K=1250$, corresponding to a a background transcription rate $k_{l}/\gamma_p=10^{-3}$.}
	\label{fig:effect.mu.t}
\end{figure}

\begin{figure}[h]
        \centering
               \subfigure[]{
\includegraphics[scale=0.5]{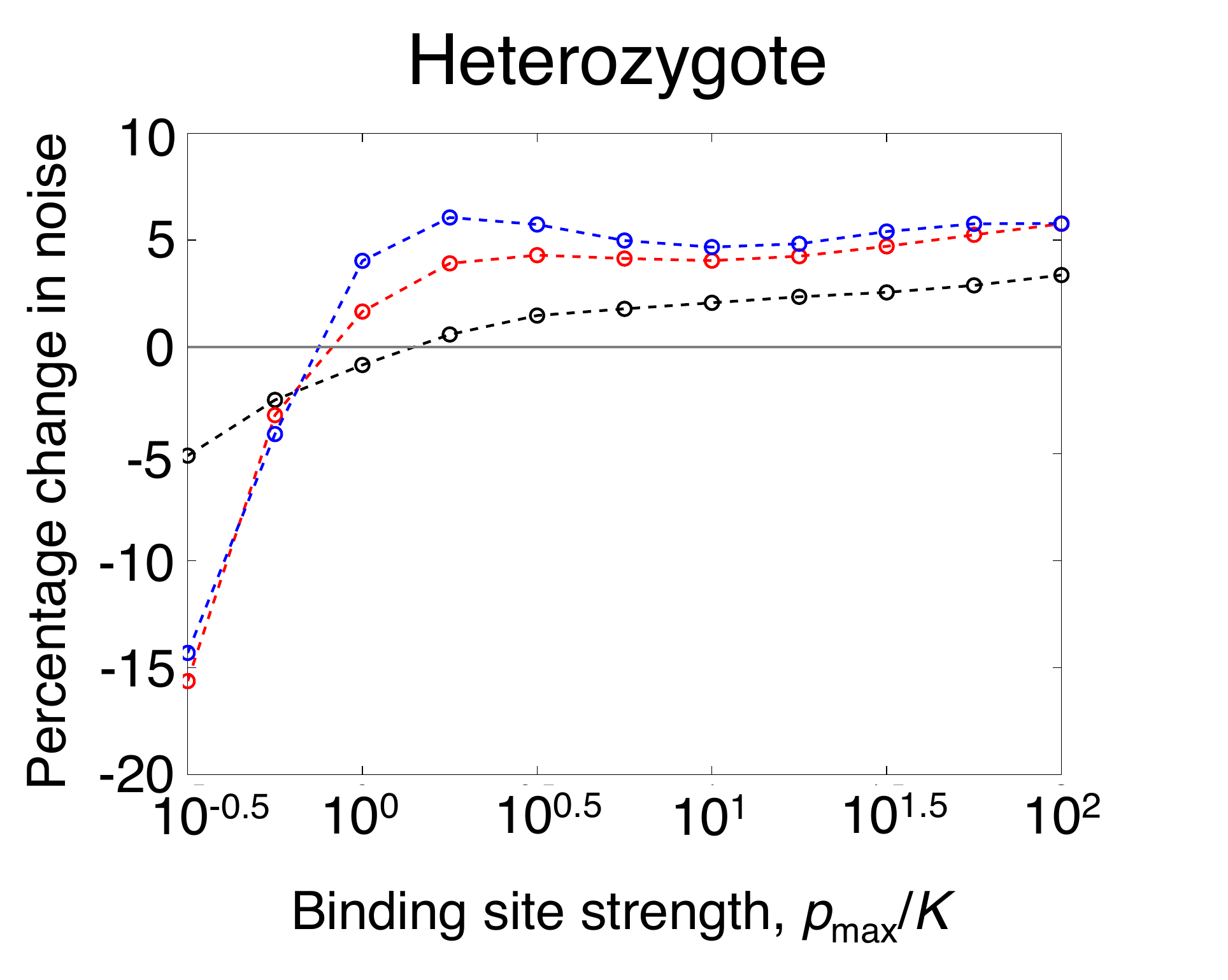} 
\label{fig:subfig1} 
} 
\subfigure[]{
\includegraphics[scale=0.5]{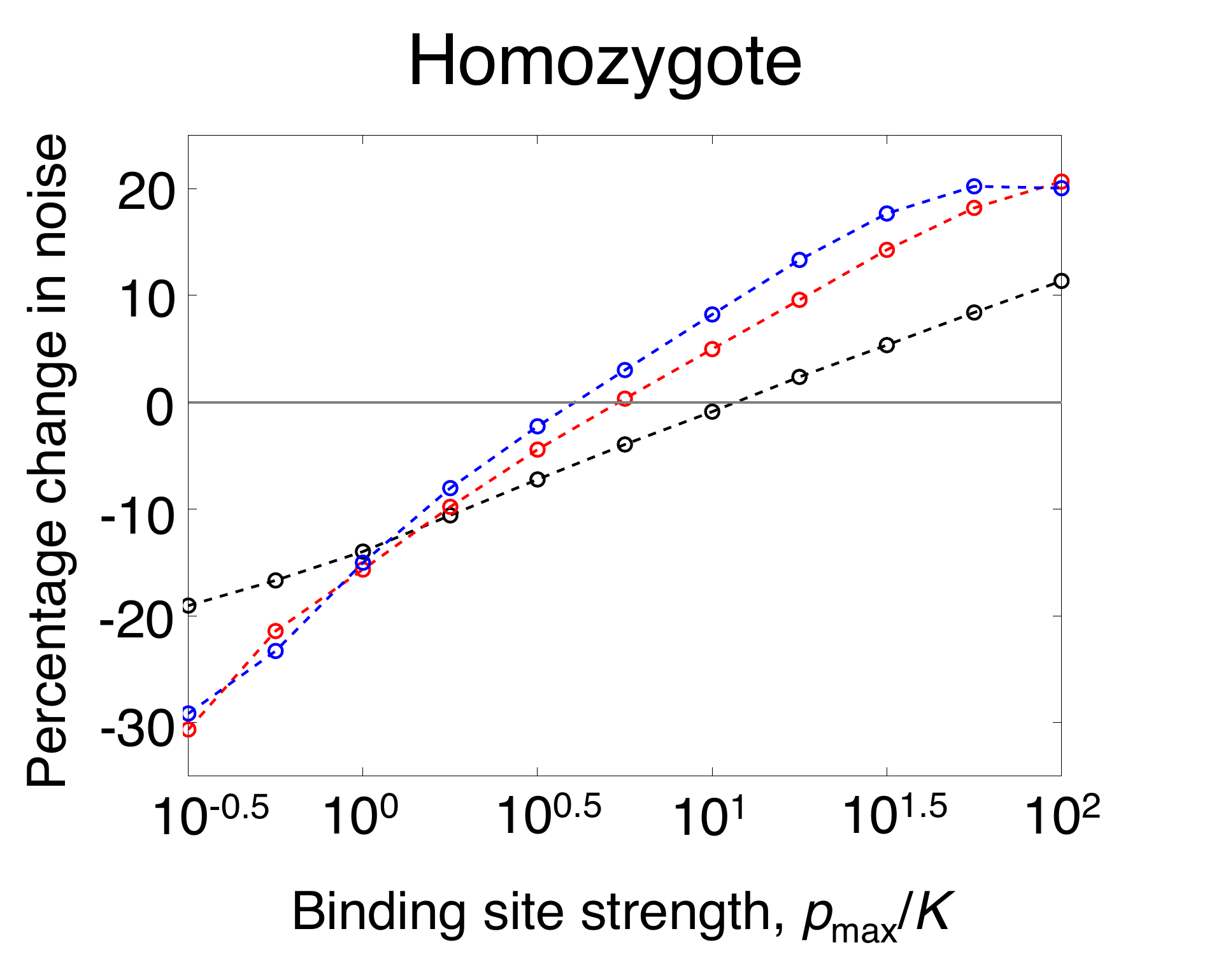}
\label{fig:subfig1} 
}
	\caption{\textbf{Intrinsic noise in gene expression}\ \ 
	The figure shows quantitative results for the intrinsic noise of autoregulating genes, as measured by the ratio of the variance to mean expression in protein concentration at equilibrium. (a) Percentage change in the noise of a heterozygote compared to the resident homozygote. These are shown for different Hill coefficients, $n=1$ (black), $n=2$ (red) and $n=3$ (blue). Mutations become deleterious in the heterozygote when $p_{max}/K>1$. (b) Percentage change in the noise of a mutant homozygote compared to the resident homozygote. Mutations become deleterious in the mutant homozygote when $p_{max}/K$ is about $10$. The graphs show the results of stochastic simulations (see Materials and Methods) for parameter values typical for transcription factors  [7], $k_{r}=0.01s^{-1}$, $k_{p}=0.17s^{-1}$, $k_{l}=0.001s^{-1}$, $\gamma_{r}=\frac{1}{120}s^{-1}$ and $\gamma_{p}=\frac{1}{3600}s^{-1}$. The resident homozygote has binding strength $p_{max}/K$ (as indicated by the x-axis), mutations are of size $\epsilon= 2$.}
\end{figure}

\clearpage

\subsection*{Intrinsic noise in diploids} 

In order to investigate the evolution of negative autoregulation as a mecahnism to reduce intrinsic noise in diploids, we turned to stochastic simulations.
Intrinsic noise in gene expression occurs because transcription and translation are inherently noisy processes.
As a result of this noise, all genes experience constant fluctuations in their mRNA and protein levels.
The greater intrinsic noise associated with a particular gene, the higher the variance in its expression level relative to the mean.
Therefore, a natural way to characterise the amount of intrinsic noise associated with a gene is to measure the ratio of the variance to the mean expression level at equilibrium (known as the Fano factor) \cite{Thattai:2001qp}.
We performed molecular simulations that capture transcription, translation and degredation in the presence of negative autoregulation (see Materials and Methods).
Just as in our analysis of response times, we compared a resident allele with dissociation constant $K_1$, to a mutant allele with dissociation constant $K_2=K_1\exp[-\epsilon]$.
We compared the intrinsic noise (as measured by the Fano factor) in the resident homozygote to that of the heterozygte and the mutant homozygote, and thus determined whether under-dominance occurs in the evolution of negative autoregulation as a mechanism to reduce intrinsic noise.
The results are shown in Fig.~5.
We find once again that under-dominance occurs.
Whereas the optimal binding strength for a single negatively autoregulating binding site is found to be $p_{max}/K \sim 10$, the maximum evolvable binding strength (i.e that which can evolve without encountering under-dominance) is found to be $p_{max}/K \sim 1$, an order of magnitude weaker.
A similar pattern occurs when steeper Hill coefficients are considered (Fig.~5).
Therefore we conclude that under-dominance poses a barrier to the evolution of strong negative autoregulation both as a mechanism to speed response times and to reduce intrinsic noise.

\subsection*{The effects of mutations to other parameters} 
To test the generality of our findings, we also considered  variation in other parameters (see SI Fig.~S1-S5 and Text). We first relaxed our assumption of a single binding site and explored the case of Hill coefficients $n>1$, implying regulation through multiple, cooperatively acting binding sites. In line with the effect of increasing binding strength through changes in $K$, we find that mutations increasing the Hill coefficient are subject to under-dominance (see SI Fig.~S1-S2 and Text). Therefore, a mutation that increases the strength of negative autoregulation are subject to the same evolutionary constraints, independent of whether they increase regulation by changing the dissociation constant $K$ or the Hill coefficient $n$.

We also considered variation in the rates of mRNA and protein degradation ($\gamma_{r}$ and $\gamma_{p}$) to see whether they provide conditions in which the effects of under-dominance on autoregulatory binding strength can be avoided (see SI Fig.~S4 and Text).
Variation in the rate of mRNA or protein degradation did not remove the tendency for mutations that increase autoregulatory binding strength to be subject to under-dominance. However, it has been pointed out in other work \cite{Eden:2011vn,Raj:2008bs}, faster rates of protein degradation result in faster response times, and regulation of protein degradation can reduce noise. As might be expected, the constraints we describe on the evolution of response times through stronger negative autoregulation do not preclude the evolution of response times through other mechanisms, such as changes in protein degradation rates.

\section*{Discussion}
Negative autoregulation is found to occur in 46\% of \textit{E. coli} transcription factors \cite{Becskei:2000vo,ABC:Lestas,Rosenfeld:2002iv,Thattai:2001qp}, but is rare in other species for which systematic data on transcriptional regulation is available, occurring in $<$2\% of the transcription factors of  yeast, \textit{Drosophila} and humans (see SI, Table S1-S3). We have argued that, at least in part, this difference can be understood by considering the different evolutionary dynamics of autoregulating genes in haploids and diploids: selection for genes to have a decreased response time to perturbations favours negative autoregulation in haploids, but under-dominance tends to prevent the evolution of stronger autoregulatory binding sites for this purpose in diploids. This constraint on the evolution of negative autoregulation in diploids is compelling because it offers a simple and general explanation for the near absence of the motif across yeast, humans and flies. Furthermore, it is important to note that under-dominance is not built into our model but arises as an emergent property of our analysis of regulatory evolution -- an analysis that simply extends to diploids previous models that have been shown to provide a good description of regulatory behaviour in haploids \cite{Rosenfeld:2002iv,Thattai:2001qp}. 

The empirical patterns we present are striking, however it is important to ask weather they can be explained by other means than those proposed in this paper.
In particular we asked whether negative autoregulation is truly under-represented in the yeast, human and \textit{Drosophila} data sets, as compared to \textit{E. coli}, or whether the apparent reduction in the number of negative autoregulators is due to under-representation of genes with repressive function generally.
To address this we interrogated each dataset to find the number of transcription factors with documented repressor activity.
These account for $58$ factors in humans, $37$ in \textit{Drosophila}, $54$ in yeast and $82$ in \textit{E. coli}.
If we include only transcription factors with known repressor function in our analysis, we find that 5 out of 58 ($8.6 \%$) genes negatively autoregulate in humans,  3 out of 37 ($8.1 \%$) in \textit{Drosophila}, 3 out of 54 ($5.6\%$) in yeast and 82 out of 138 ($59\%$) in \textit{E. coli}.
Thus, the relative rarity of negative autoregulation in eukaryotes is not due to a general underrepresentation of repressive transcription factor effects among the genetic interactions described for these species. Instead, they appear to be a true property of their regulatory networks.

The species for which we were able to collate data show stark differences in the frequency of negative autoregulation. However, it would clearly be desirable to extend the scope of the empirical analysis to include examples that span the prokaryote-eukaryote divide. Two types of data are of particular interest in this respect: haploid genes in diploid species and duplicate genes in haploid species. The constraints on the evolution of autoregulation predicted by our model only affect genes for which two homologous copies are expressed in the same cell. Haploid genes in a diploid organism should therefore escape the evolutionary constraint on negative autoregulation. Unfortunately, the data on genetic regulation are currently too sparse to test this prediction with any degree of rigor. The only candidate for a haploid gene in our dataset is the human Y-linked transcription factor \textit{Sry} (see SI Table.~S3). However, its mode of regulation (positive or negative) is unknown. Duplicate genes in haploids offer a better prospect. Just as single copy genes in diploids are predicted to escape under-dominance, multi-copy genes in haploids may be subject to evolutionary constraints similar to those we have described for diploids. Interestingly, this appears to be supported by data on duplicate genes in \textit{E. coli}. Although the use of negative autoregulation is widespread amongst \textit{E. coli} transcription factors, duplicates of negative autoregulating genes are no more common than among other genes \cite{Warnecke:2009fi}. This is despite the fact that negative autoregulation is expected to reduce the deleterious effects of increased dosage following duplication \cite{Warnecke:2009fi}. Although the evolutionary dynamics of duplication and divergence are complex \cite{Innan:2010ly}, it is interesting to note that our model predicts  that any divergence in the expression levels of a pair of negatively autoregulating duplicate genes will tend to slow the response time of the pair. This is because expression divergence will tend to increase the difference in expression across the two genes and thus will increase the response time in exactly the same way as we have described for heterozygotes in diploid cells.

The frequency of autoregulating duplicates in \textit{E. coli} offers some support for our model by showing that, as we would predict, the evolution of multi-copy genes in bacteria is constrained. However, it could be argued that the relative dearth of negative autoregulation in yeast, fruit flies and humans is not primarily caused by under-dominance, but is rather due to the fact that these organisms are eukaryotes. Eukaryotes may simply experience different types of noise, and, accordingly, have different mechanisms for dealing with it, making negative autoregulation unnecessary. Although we cannot dismiss this limitation of our model entirely, several points are worth noting. The use of response time as a measure of fitness makes our model very general. First, any cell has to deal with large perturbations, such as occur following cell division. The speed with which the concentration of a transcription factor returns to its equilibrium, and the regulatory dynamics allowing it to do so, are important across all levels of biological complexity. Second, as we have noted, although our model captures the response time to perturbations and the amount of intrinsic noise associated with a gene \cite{Thattai:2001qp}, it does not capture other, extrinsic sources of noise. In particular, eukaryotes tend to be affected by ``input noise'' that results, for example, from the stochastic ON-OFF switching occuring in eukaryotic cells \cite{ABC:Raj2006,Raj:2008bs}, and it is positive autoregulation, not negative autoregulation, that is best for dealing with such noise \cite{Gregor:2007fk,ABC:Raj2006,Raj:2008bs}. However, positive autoregulation does not feature any more prominently than negative autoregulation within the regulatory networks of the three eukaryotes we analysed, with 9 instances in yeast, 16 in humans and 11 in \textit{Drosophila}. These figures are not comparable to the frequency of negative autoregulation in \textit{E. coli}, indicating that we are not simply observing a straightforward shift in the importance of different types of perturbations.


It seems unlikely that eukaryotes are completely exempt from dealing with the kind of perturbations that would require negative autoregulation. It is possible, however, that they implement autoregulation differently, and in a way that would not appear in our systematic screen of existing datasets. When using transcription regulation, eukaryotes may be able to
achieve some degree of negative autoregulation through multiple, weak autoregulatory binding sites,
along with cooperation (see Figs.~ S1-S2). Although we are able to show that the evolution of strong cooperative autoregulation is subject to under-dominance (Fig.~S1), we find that the evolution of multiple, weak autoregulatory binding sites (Fig.~ S2) are less constrained. Since weak binding sites would likely be under-represented or absent from systematic datasets, we cannot rule out the possibility that diploids achieve negative autoregulation in this way, and some studies based on sequence conservation do suggest that autoregulatory binding sites in humans may be quite widespread \cite{Kiebasa:2008kx}.
 Eukaryotes may also achieve negative autoregulation through mechanisms other than direct transcription regulation, for example, through changes in local chromatin structure or covalent changes in the protein structure of transcription factors. As these regulatory mechanisms are less likely to generate the phenomena of cross-regulation that occur in diploid transcription regulation, they may not to be subject to under-dominance.
Finally, it is important to reiterate that our study is only concerned with the evolution of negative autoregulation for noise reduction and faster response times. However, genes can achieve noise reduction through other means than autoregulation, and autoregulation can be used for other purposes than noise reduction \cite{Amit:2007fk,Legewie:2008uq,Schwanhausser:2011vn}. We do not suggest that eukaryotes are exempt from the problem of noise. We do suggest that diploid gene networks, in contrast to those of haploids, must seek a different solution to the same problem.

\subsection*{Conclusion} Our work shows that regulatory interactions between homologous genes can generate deleterious effects that constrain the evolution of negative autoregulation. The predictions of our model show that the high incidence of autoregulation in \textit{E. coli} and the dearth of negatively autoregulating genes in yeast, flies and humans can be reconciled by taking into account a simple biological attribute---ploidy. Importantly, the difference between haploid and diploid regulation is not a mere correlate of the prokaryote-eukaryote divide. This was already suggested by the finding that the genetic networks of \textit{E. coli} and yeast are---with the exception of their use of autoregulation--- very similar \cite{Milo:2004ez} and is further corroborated by the the fact that the predictions of our model are met by data on single and duplicated genes within \textit{E.~coli}.

More generally, our work demonstrates that regulatory evolution can be considerably complicated by the presence of multiple copies of a gene in a cell, as is typically the case for eukaryotes. By explicitly considering the evolution of regulatory interactions, we have highlighted constraints that would not be evident from an analysis of the functional properties of an existing regulatory interaction in isolation---strong negative autoregulation quickens the response of genes to perturbation, but it is hard to evolve for this purpose due to under-dominance. This evolutionary perspective needs to be absorbed into attempts at unravelling the function of regulatory networks in higher organisms, a key problem for systems biology.

\section*{Methods}

\subsection*{Monte-Carlo simulations}
\noindent We used simulations of the molecular dynamics within a cell to determine the amout of intrinsic noise of autoregulating genes in diploids. 
A model that tracks the number of mRNA and protein molecules for a negatively autoregulating gene within a haploid cell is described in \cite{Thattai:2001qp}. We generalised this to account for diploidy.
The state of the system is described by the number of mRNA molecules $r_i$, and the number of protein molecules $p_i$ produced from the two alleles $i \in \{{1,2}\}$.
The probability of a state $\{r_1,r_2,p_1,p_2\}$ is specified by the joint probability distribution $n_{r_1,r_2,p_1,p_2}(t)$. The transition probabilities for the system to move between states due to changes in $r_1$ and $p_1$ (and, analogously, due to changes in $r_2$ and $p_2$) are given by
\begin{eqnarray}
\nonumber \{r_1,r_2,p_1,p_2\}&\xrightarrow{\ k_l +\phi_1(p)\ }&\{r_1+1,r_2,p_1,p_2\} ,\\
\nonumber \{r_1,r_2,p_1,p_2\}&\xrightarrow{\: r_1k_{p}}&\{r_1,r_2,p_1+1,p_2\} ,\\
\nonumber \{r_1,r_2,p_1,p_2\}&\xrightarrow{r_1\gamma_{r}}&\{r_1-1,r_2,p_1,p_2\} ,\\
\nonumber \{r_1,r_2,p_1,p_2\}&\xrightarrow{\ p_1\gamma_{p}\; }&\{r_1,r_2,p_1-1,p_2\} ,
\end{eqnarray}
\\
where $p=p_1+p_2$, $k_l+\phi_1(p)$ is the rate at which mRNA molecules are transcribed from allele 1, $\gamma_{r}$ is the rate of mRNA degradation, $k_{p}$ is the rate at which mRNA is translated into protein and $\gamma_{p}$ is the rate of protein degradation.
As in the ODE model, $\phi_1(p)$ is a function of the number of proteins $p$ present in the cell, such that
\begin{equation*}
\phi_1(p)=\frac{k_{0}}{1+\frac{p}{K_1}} ,
\end{equation*}
\\
where $k_{0}$ is the maximum rate of mRNA transcription, and $K_1$ is the dissociation constant of the binding site of allele 1.

To calculate response times we first determined the equilibrium expression level of the system from the average of $10^5$ replicate Monte-Carlo simulations. We then reduced mRNA and protein levels to a fraction $\alpha$ of the equilibrium level. The time for each replicate to return to equilibrium was measured and the average across the ensemble used as an estimate of the response time of the system. In order to determine how response times vary with the level of perturbation, simulations were run for values of $\alpha$ between 0 and 1 in steps of 0.01.

\subsection*{Simulations of binding site evolution}
\noindent Binding site evolution was modelled by generating a transcription factor binding motif with a length $n$ nucleotides and an optimal base associated with each nucleotide. As in other models of TF-DNA binding, when a given nucleotide $i$ was matched for for the optimal base it contributed an amount $\epsilon_{i}$ to binding energy, otherwise it contributed 0 \cite{Gerland:2002cv,Lassig:2007fk,Berg:2004uq}.

\noindent Binding site lengths were drawn from an empirical distribution generated from the binding motifs of 454 eukaryotic transcription factors contained in the JASPAR CORE database \cite{Bryne:2008fk}. 
The value of $\epsilon_i$ for each nucleotide was drawn from a uniform distribution in the interval $(0,3)$. 
The optimal binding strength $K_{opt}$ was determined numerically (see Appendix), using the values for the system parameters that are given in the legend of Figure 4.
We excluded from our analysis any binding sites for which the total binding strength of the optimal sequence was too low to achieve the fastest response time (i.e., those sequences for which $K=\exp[-\sum_i \epsilon_i]>K_{opt}$). Evolution started from a state of minimum affinity (all nucleotides non-optimal) and proceeded through a series of single nucleotide substitutions. At each time step, a random mutation was introduced into the binding site sequence, switching one nucleotide from the non-optimal to the optimal state. If the mutation resulted in a response time less than or equal response time of the resident, the mutant sequence was assumed to go to fixation in the population. Deleterious mutations that increased response times were assumed to be lost. The simulation was ended when no further advantageous mutations were available. Simulations were carried out for both haploids and diploids (for which response time of mutants was evaluated in the heterozygote state).

\subsection*{Derivation of response times in haploids}

\noindent Here we derive results for the response time of a haploid autoregulating gene. We derive results for the general case in which autoregulation is described by a Hill function with arbitrary coefficient $n$ (the analyses in the main text assumes $n=1$).

The set of ODEs describing transcription and translation of mRNA and protein at a single autoregulating gene are analogous to those given for one allele in Eqs.~1 for a pair of autoregulating genes in a diploid. In order to simplify the analysis of the system we make the change of variables
\begin{eqnarray*}
\nonumber s&=&\frac{r}{r_{max}-r_{min}} ,\\
\nonumber q&=&\frac{p}{p_{max}-p_{min}} ,\\
\tau&=&\gamma_{p}t ,
\end{eqnarray*}
\\
with $L=\frac{K}{p_{max}-p_{min}}$ and $\gamma=\frac{\gamma_{p}}{\gamma_{r}}$.
The dynamics of the system can then be rewritten as
\begin{eqnarray}
\nonumber \gamma\frac{ds}{d\tau}&=&\beta + k_{s}(q)-s ,\\
\frac{dq}{d\tau}&=&s-q ,
\end{eqnarray}
\\
where $\beta=\frac{r_{min}}{r_{max}-r_{min}}=\frac{p_{min}}{p_{max}-p_{min}}=\frac{k_{l}}{k_{0}}$. In general $\beta\ll1$ since $k_{l}\ll k_{0}$ and
\begin{equation}
k_{s}(q)=\frac{1}{1+\left(\frac{q}{L}\right)^{n}} \hspace{0.2cm},
\end{equation}
\\ 
is the rescaled form of the repression function $\phi(p)$ described in the main text.
Assuming that mRNA decays much faster than protein \cite{Rosenfeld:2002iv,Thattai:2001qp} $\gamma_{p}\ll\gamma_{r}$, then  $\gamma\ll1$, it follows that $\gamma\frac{ds}{d\tau}$ is small relative to $\frac{dq}{d\tau}$ and we can assume that transcription output goes to equilibrium rapidly. That is, we can take $\gamma\frac{ds}{d\tau}\approx0$  and hence that the quasi equilibrium condition $s\approx \beta + k_{s}(q)$ holds. Substituting into Eqs.~3, generates a 2-dimensional system that is well approximated by the 1-dimensional system
\begin{equation}
\frac{dq}{d\tau}=\beta+k_{s}(q)-q .
\end{equation}

\subsection*{Small Perturbations}
The Lyapunov exponent associated with Eq.~5 at equilibrium gives the rate at which the system returns to equilibrium following a small perturbation.
It is given by
\begin{equation}
\lambda=-\left(1+\left|\frac{dk_{s}}{dq}\right|\right) .
\end{equation}
\\
Eq. 6 is always negative. In what follows we will discuss only the magnitude of the Lyapunov exponent $\left|\lambda\right|$ with the understanding that this quantity is always negative and therefore describes the rate at which the system returns to equilibrium.  From Eq.~6 it is clear that a mutation which increases $\left|\frac{dk_{s}}{dq}\right|$ will always serve to decrease the Lyapunov exponent and thus increase the rate at which the system converges to equilibrium. 

\subsection*{Evolution of a new binding site}
We compare a wild-type binding site, with dissociation constant $L_{1}$, to a mutant binding site with dissociation constant $L_{2}$ such that $L_{1}>L_{2}$ ---meaning that the mutant has a stronger binding site than the wild-type. At equilibrium, the protein concentrations satisfy
\begin{eqnarray}
\nonumber q^{*}_{1}&=&\beta+\frac{1}{1+\left(\frac{q^{*}_{1}}{L_{1}}\right)^{n}} \hspace{0.2cm},\\
q^{*}_{2}&=&\beta+\frac{1}{1+\left(\frac{q^{*}_{2}}{L_{2}}\right)^{n}} \hspace{0.2cm}.
\end{eqnarray}
\\
It is simple to show that $q^{*}_{1}>q^{*}_{2}$ by differentiating, with respect to $L$.
Thus, strengthening the autoregulatory binding site (i.e., decreasing $L$) will lead to a decrease in the equilibrium protein concentration, and so with $L_{1}>L_{2}$ we always have $q^{*}_{1}>q^{*}_{2}$.
To calculate the value of $L_{opt}$ for which $\left|\lambda\right|$ is maximum, we note that
\begin{equation*}
\left|\frac{dk_{s}}{dq}\right|=\frac{n}{q}k_{s}(q)(1-k_{s}(q)) .
\end{equation*}
\\
At equilibrium $q=q^*=k_{s}(q^*)+\beta$, and the Lyapunov exponent can be written as
\begin{equation*}
\left|\lambda\right|=\left(1+\frac{n}{q^*}(q^*-\beta)(1-q^*+\beta)\right) ,
\end{equation*}
\\
and we can find the value of $q^*$ that results in the largest Lyapunov exponent. This is given by
\begin{equation*}
q^*=\sqrt{\beta(1+\beta)} \hspace{0.2cm}.
\end{equation*}
\\
Translating this back into units of protein concentration, this means that the fastest response to small perturbations about equilibrium occurs when
\begin{equation}
p^*=\sqrt{p_{max}p_{min}} \hspace{0.2cm}.
\end{equation}
\\
Thus, mutations which increase the strength of negative autoreguation, (and therefore decrease $K$), will decrease response time provided the equilibrium protein concentration is $p^*>\sqrt{p_{max}p_{min}}$, as discussed in the main text. The optimal binding site strength $K_{opt}$ can be determined by calculating the value of $K$ which gives the optimal equilibrium protein concentration of Eq.~8. In the general case of arbitrary $n$, $K_{opt}$ cannot be found analytically, but it can always be found numerically.

The derivation of $K_{opt}$ presented here is based on the assumption that perturbations of the system are small, in which case the dynamics of the system are well captured by its Lyapunov exponent. The optimal binding strength under perturbations of arbitrary size can be obtained by numerical integration of the system. As might be expected, the values$K_{opt}$ obtained in this way are similar to those calculated for small perturbations above.

\subsection*{Derivation of response times in diploids}

\noindent For diploids we proceed in the same way as for a single gene, and obtain a 1-D system for expression of a pair of alleles (with dissociation coefficients $L_i$ and $L_j$) 
\begin{equation}
\frac{dq_{ij}}{d\tau}=2\beta+k_{s_{i}}(q_{ij})+k_{s_{j}}(q_{ij})-q_{ij} \hspace{0.2cm},
\end{equation}
\\
where
\begin{equation*}
k_{s_{i}}(q_{ij})=\frac{1}{1+\left(\frac{q_{ij}}{L_{i}}\right)^{n}} \hspace{0.2cm}.
\end{equation*}

\subsection*{Evolution of a new binding site}
We now consider the response time of a pair of autoregulating alleles in a diploid.  When an organism is  homozygous, both binding sites have the same dissociation constant, $L_{1}$ and Eq.~9 is of the same form as Eq.~5 for a haploid, and the results for response time in haploids can be applied. When an organism is heterozygous however, the results for haploids do not hold. We compare the Lyapunov exponents of a heterozygote with dissociation constants $L_1$ and $L_2$, where $L_{2}<L_{1}$, to a resident homozygote in which both binding sites have strength $L_{1}$. At equilibrium the total protein concentrations satisfy
\begin{eqnarray}
\nonumber q^{*}_{11}&=&2\beta+\frac{2}{1+\left(\frac{q^{*}_{11}}{L_{1}}\right)^{n}} \hspace{0.2cm},\\
q^{*}_{12}&=&2\beta+\frac{1}{1+\left(\frac{q^{*}_{12}}{L_{1}}\right)^{n}}+\frac{1}{1+\left(\frac{q^{*}_{12}}{L_{2}}\right)^{n}} \hspace{0.2cm},
\end{eqnarray}
\\
where $q_{11}^{*}$ is the equilibrium protein concentration of the (resident) homozygote and $q_{12}^{*}$ is the equilibrium expression of the (mutant) heterozygote.
It is simple to show that $q_{11}^{*}>q_{12}^{*}$. by differentiating Eq.~10 with respect to $L_{2}$.

\subsection*{Small perturbations}
Following a small displacement from equilibrium, under-dominance will occur if the heterozygote has a smaller Lyapunov exponent than the homozygote. The maximal Lyapunov exponent of the system is given by
\begin{equation}
\left|\lambda_{11}\right|=1+\frac{2n}{q_{11}^{*}}\left(\frac{q_{11}^*}{2}-\beta\right)\left(1-\frac{q_{11}^*}{2}+\beta\right) ,
\end{equation} 
\\
for the homozygote, and
\begin{eqnarray}
\nonumber \left|\lambda_{12}\right|=&1+\frac{n}{q_{12}^{*}}\left(q_{1,12}^*-\beta\right)\left(1-q_{1,12}^*+\beta\right)\\
&+\frac{n}{q_{12}^{*}}\left(q_{2,12}^*-\beta\right)\left(1-q_{2,12}^*+\beta\right) ,
\end{eqnarray}
\\
for the heterozygote, where $q_{i,ij}$ referes to allele $i$ in a diploid carrying alleles $i$ and $j$. We can observe that the squared difference in the mean allele expression, $V$, is given by $V=\left(q_{1,12}^*-\frac{q_{12}^*}{2}\right)^2+\left(q_{2,12}^*-\frac{q_{12}^*}{2}\right)^2$, which can be expanded to give
\begin{equation*}
V=(q_{1,12}^*)^2+(q_{2,12}^*)^2-\frac{q_{12}^*}{2} .
\end{equation*}
\\
Substituting this expression for $V$ in Eq. 12 we find
\begin{eqnarray}
\left|\lambda_{11}\right|&=&1+n\left(1-\frac{q_{11}^*}{2}+2\beta-\frac{2\beta}{q_{11}^*}(1+\beta)\right) ,\\
\left|\lambda_{12}\right|&=&1+n\left(1-\frac{q_{12}^*}{2}+2\beta-\frac{2\beta}{q_{12}^*}(1+\beta)-\frac{V}{q_{12}^*}\right) .
\end{eqnarray}
\\
Note that Eq.~14 is of the same form as Eq.~13, with an additional term that depends on the ratio of the squared difference in allele expression, $V$ to the total expression.
We can define $\lambda_{hom}$ to be the Lyapunov exponent associated with a homozygote of a given equilibrium expression and  $\lambda_{het}$  to be the Lyapunov exponent associated with a heterozygote of the same equilibrium expression and obtain Eq.~2 of the main text (with $n=1$).



\section*{Acknowledgments}
The authors thank Joshua Plotkin for a great deal of advice and feedback. 


\end{document}